\documentclass{Interspeech}

\usepackage[acronym]{glossaries}
\usepackage{microtype}

\glsdisablehyper

\interspeechcameraready 

\newacronym{bsrnn}{BSRNN}{band-split recurrent neural network}
\newacronym{casp}{CASP}{computational model of human auditory signal processing and perception}
\newacronym{drnl}{DRNL}{dual-resonance non-linear}
\newacronym{erb}{ERB}{equivalent rectangular bandwidth}
\newacronym{estoi}{ESTOI}{extended short-term objective intelligibility}
\newacronym{fir}{FIR}{finite impulse response}
\newacronym{gmac}{GMAC}{giga multiply-accumulate}
\newacronym{gpu}{GPU}{graphics processing unit}
\newacronym{haspi}{HASPI}{hearing aid speech perception index}
\newacronym{hasqi}{HASQI}{hearing aid speech quality index}
\newacronym{hi}{HI}{hearing impaired}
\newacronym{hlc}{HLC}{hearing loss compensation}
\newacronym{lstm}{LSTM}{long short-term memory}
\newacronym{mae}{MAE}{mean absolute error}
\newacronym{mlp}{MLP}{multilayer perceptron}
\newacronym{mse}{MSE}{mean squared error}
\newacronym{nh}{NH}{normal hearing}
\newacronym{nr}{NR}{noise reduction}
\newacronym{pesq}{PESQ}{perceptual evaluation of speech quality}
\newacronym{snr}{SNR}{signal-to-noise ratio}
\newacronym{sdr}{SDR}{signal-to-distortion ratio}
\newacronym{stft}{STFT}{short-time Fourier transform}
\newacronym{rir}{RIR}{room impulse response}

\mathchardef\mhyphen="2D

\DeclareSIUnit{\nothing}{\relax}

\makeatletter
\def\bstctlcite{\@ifnextchar[{\@bstctlcite}{\@bstctlcite[@auxout]}}
\def\@bstctlcite[#1]#2{\@bsphack
  \@for\@citeb:=#2\do{%
    \edef\@citeb{\expandafter\@firstofone\@citeb}%
    \if@filesw\immediate\write\csname #1\endcsname{\string\citation{\@citeb}}\fi}%
  \@esphack}
\makeatother

\makeatletter
\def\blfootnote{\gdef\@thefnmark{$\dagger$}\@footnotetext}
\makeatother

\title{Controllable joint noise reduction and hearing loss compensation using a differentiable auditory model}

\author[affiliation={1,\dagger}]{Philippe}{Gonzalez}
\author[affiliation={2}]{Torsten}{Dau}
\author[affiliation={2}]{Tobias}{May}

\affiliation{}{Demant A/S}{Denmark}
\affiliation{}{Department of Health Technology}{Technical University of Denmark}

\email{phgo@demant.com, tdau@dtu.dk, tobmay@dtu.dk}


\begin{document}

\bstctlcite{IEEEexample:BSTcontrol}

\maketitle

{
\makeatletter\def\Hy@Warning#1{}\makeatother  %
\blfootnote{Work done while at Technical University of Denmark.}
}

\begin{abstract}
Deep learning-based \gls{hlc} seeks to enhance speech intelligibility and quality for hearing impaired listeners using neural networks.
One major challenge of \gls{hlc} is the lack of a ground-truth target.
Recent works have used neural networks to emulate non-differentiable auditory peripheral models in closed-loop frameworks, but this approach lacks flexibility.
Alternatively, differentiable auditory models allow direct optimization, yet previous studies focused on individual listener profiles, or joint \gls{nr} and \gls{hlc} without balancing each task.
This work formulates \gls{nr} and \gls{hlc} as a multi-task learning problem, training a system to simultaneously predict denoised and compensated signals from noisy speech and audiograms using a differentiable auditory model.
Results show the system achieves similar objective metric performance to systems trained for each task separately, while being able to adjust the balance between \gls{nr} and \gls{hlc} during inference.
\end{abstract}

\section{Introduction}

Hearing impairment affects millions of people worldwide, impacting social interactions, cognitive function, and overall quality of life.
While hearing aids can mitigate these challenges, users often report difficulties understanding speech in complex acoustic environments.
Currently, \gls{nr} and \gls{hlc} in hearing aids primarily rely on simplistic algorithms like beamforming and frequency band amplification.
However, deep learning offers the potential to surpass these traditional methods, as it allows complex non-linear mappings between noisy speech, listener profiles, and optimal \gls{hlc} strategies.

Unlike conventional tasks such as speech enhancement where defining a training objective using the ground-truth target is straightforward, \gls{hlc} is more challenging due to the absence of such a target.
Recent studies have tackled this by training auxiliary neural networks to emulate non-differentiable auditory peripheral models~\cite{baby2021convolutional,drakopoulos2021convolutional,leer2024train}.
These emulators were then used in closed-loop optimization frameworks~\cite{drakopoulos2022differentiable,drakopoulos2023neural,drakopoulos2023dnn,leer2025hearing}.
While this approach facilitates \gls{hlc} algorithm training, it lacks flexibility, since retraining is required whenever the auditory model is updated.
Additionally, most studies have trained these systems for individual listener profiles, using auditory models describing detailed physiological processes, some of which may not be necessary for achieving perceptual benefits in hearing aids.

Another approach involves designing differentiable auditory models that can directly be integrated into the optimization of the \gls{hlc} algorithm.
This makes it easier to explore which auditory model stages are essential for delivering perceptual benefits.
Studies adopting this approach~\cite{tu2021dhasp,tu2021optimising,tu2021two} have shown that trained \gls{hlc} algorithms can outperform traditional hearing aid prescriptions in both quiet~\cite{tu2021dhasp} and noisy~\cite{tu2021optimising} conditions, as measured by objective metrics.
However, these efforts often utilize simplistic systems with very few trainable parameters, such as fixed \gls{fir} filterbanks with learnable gains, and they continue to train systems for individual listener profiles.

Recent studies have trained listener-independent systems for \gls{nr} and \gls{hlc}~\cite{zmolikova2021but,cheng2023speech,drgas2024dynamic}.
However, the proposed algorithms are unable to perform \gls{nr} or \gls{hlc} in a controllable manner during inference.
Studies have shown that there exist sub-populations of HI listeners who prefer strong \gls{nr} over mild \gls{nr}~\cite{neher2014relating,neher2016directional}.
Additionally, users may want to prioritize \gls{nr} or \gls{hlc} depending on the listening environment.
For example, in very noisy environments, users may wish for strong \gls{nr}, while in social gatherings, they may prefer \gls{hlc} without \gls{nr}.
Therefore, the ability to adjust the balance between \gls{nr} and \gls{hlc} during inference can be a valuable feature in hearing aids.

In this work, we propose a multi-task learning framework for joint \gls{nr} and \gls{hlc}.
A differentiable auditory model is used for training a speech processor to simultaneously predict a denoised and a compensated signal from a noisy input speech signal and an audiogram.
Each task is assigned a distinct training objective, and these objectives are combined using an uncertainty-based weighting scheme~\cite{kendall2018multi}.
During inference, the system can flexibly mix the two output signals to enable controllable joint \gls{nr} and \gls{hlc}.
Code is available online at \url{https://github.com/philgzl/cnrhlc}.

\section{Typical framework}
\label{sec:typical}

\begin{figure}
  \centering
  \includegraphics{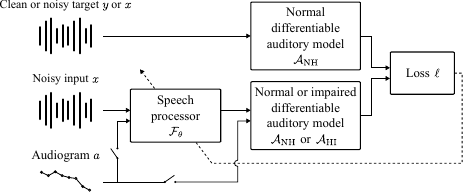}
  \caption{Typical framework for \gls{nr}-only, \gls{hlc}-only, or joint \gls{nr} and \gls{hlc} without control.}
  \label{fig:framework}
\end{figure}

Figure~\ref{fig:framework} shows a typical framework for training a speech processor~$\mathcal{F}_\theta$ for \gls{nr}-only, \gls{hlc}-only, or joint \gls{nr} and \gls{hlc} without control.
The speech processor input is noisy speech~$x$ and its output is fed to a \gls{nh} or \gls{hi} differentiable auditory model~$\mathcal{A}_\mathrm{NH}$ or $\mathcal{A}_\mathrm{HI}$.
If the task includes \gls{hlc}, $\mathcal{F}_\theta$ and $\mathcal{A}_\mathrm{HI}$ also take an audiogram~$a$ as input.
The target signal is the corresponding clean speech~$y$ or the same noisy speech~$x$, and is fed to a \gls{nh} auditory model~$\mathcal{A}_\mathrm{NH}$.
The speech processor is then optimized to minimize a loss function~$\ell$ between the output of the two auditory models.
Whether the auditory model at the output of the speech processor is \gls{nh} or \gls{hi} and the target signal is clean or noisy depends on the task:
\begin{itemize}
  \item If the auditory model at the output of the speech processor is \gls{nh} and the target signal is clean, then the speech processor is optimized for \gls{nr}-only.
  The training objective is
  \begin{equation}
    \label{eq:loss_nr}
    \mathcal{L}_\mathrm{NR} = \ell \Bigl( \mathcal{A}_\mathrm{NH} \bigl( \mathcal{F}_\theta(x) \bigr), \mathcal{A}_\mathrm{NH}(y) \Bigr).
  \end{equation}
  This is similar to traditional speech enhancement, except that the training objective is based on an auditory model instead of e.g.\ \gls{sdr}.
  For this task, no audiogram is fed to the speech processor nor the auditory model at its output.
  While denoising can improve the intelligibility and quality of speech, \gls{hi} listeners require additional \gls{hlc}.
  \item If the auditory model at the output of the speech processor is \gls{hi} and the target signal is noisy, then the speech processor is optimized for \gls{hlc}-only,
  \begin{equation}
    \label{eq:loss_hlc}
    \mathcal{L}_\mathrm{HLC} = \ell \Bigl( \mathcal{A}_\mathrm{HI} \bigl( \mathcal{F}_\theta(x, a), a \bigr), \mathcal{A}_\mathrm{NH}(x) \Bigr).
  \end{equation}
  The speech processor is tasked with compensating for the hearing loss modeled in the auditory model at its output.
  However, background noise is not removed, which can be detrimental to the intelligibility and quality of speech.
  \item If the auditory model at the output of the speech processor is \gls{hi} and the target signal is clean, then the speech processor is optimized for joint \gls{nr} and \gls{hlc},
  \begin{equation}
    \label{eq:loss_nr_hlc}
    \mathcal{L}_\mathrm{NR \mhyphen HLC} = \ell \Bigl( \mathcal{A}_\mathrm{HI} \bigl( \mathcal{F}_\theta(x, a), a \bigr), \mathcal{A}_\mathrm{NH}(y) \Bigr).
  \end{equation}
  However, since a single loss term is used, it is unclear if the speech processor prioritizes \gls{nr} or \gls{hlc}, and it is not possible to control for each task at inference time.
\end{itemize}

\section{Proposed framework}
\label{sec:proposed}

\begin{figure}
  \centering
  \includegraphics{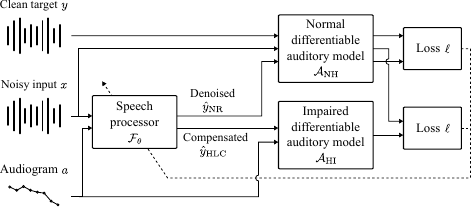}
  \caption{Proposed framework for controllable joint \gls{nr} and \gls{hlc}.}
  \label{fig:proposed}
\end{figure}

Figure~\ref{fig:proposed} shows the proposed framework for training the speech processor~$\mathcal{F}_\theta$ for controllable joint \gls{nr} and \gls{hlc}.
Given noisy speech~$x$ and an audiogram~$a$, the speech processor outputs both a denoised signal~$\hat{y}_\mathrm{NR}$ and a compensated signal~$\hat{y}_\mathrm{HLC}$,
\begin{equation}
  \mathcal{F}_\theta(x, a) = (\hat{y}_\mathrm{NR}, \hat{y}_\mathrm{HLC}).
\end{equation}
A loss term is defined for each task,
\begin{align}
  \mathcal{L}_\mathrm{NR} &= \ell \bigl( \mathcal{A}_\mathrm{NH} ( \hat{y}_\mathrm{NR} ), \mathcal{A}_\mathrm{NH}(y) \bigr), \\
  \mathcal{L}_\mathrm{HLC} &= \ell \bigl( \mathcal{A}_\mathrm{HI} ( \hat{y}_\mathrm{HLC}, a ), \mathcal{A}_\mathrm{NH}(x) \bigr).
\end{align}
The final training objective can be defined as a weighted sum of $\mathcal{L}_\mathrm{NR}$ and $\mathcal{L}_\mathrm{HLC}$.
However, finding the optimal weights using a grid search is time-consuming.
Moreover, the results can be very sensitive to the choice of the weights, and the optimal weights can vary during training.
One option is to model the predictions as isotropic Gaussian distributions, and adjust the weights dynamically based on the uncertainty of the predictions~\cite{kendall2018multi}.
In practice, the method consists in optimizing two additional parameters ${u_\mathrm{NR} = \log \sigma_\mathrm{NR}^2}$ and ${u_\mathrm{HLC} = \log \sigma_\mathrm{HLC}^2}$, where $\sigma_\mathrm{NR}>0$ and $\sigma_\mathrm{HLC}>0$ represent the homoscedastic uncertainty related to each task.
The final training objective is
\begin{equation}
  \label{eq:loss_c_nr_hlc}
  \mathcal{L}_\mathrm{C \mhyphen NR \mhyphen HLC} = \frac{\mathcal{L}_\mathrm{NR}}{e^{u_\mathrm{NR}}} + u_\mathrm{NR} + \frac{\mathcal{L}_\mathrm{HLC}}{e^{u_\mathrm{HLC}}} + u_\mathrm{HLC}.
\end{equation}
Intuitively, each loss term is weighted down if the related uncertainty is high.
At the same time, since the log-uncertainties are added to the final training objective, they are encouraged to be as small as possible.

Similar to how unprocessed and beamformed signals are mixed in hearing devices~\cite{gossling2020performance}, the denoised and compensated signals are mixed using a parameter~$\alpha \in [0, 1]$,
\begin{equation}
  \hat{y} = \alpha \hat{y}_\mathrm{NR} + (1 - \alpha) \hat{y}_\mathrm{HLC}.
\end{equation}
This allows the balance between \gls{nr} and \gls{hlc} to be adjusted without the need to retrain the speech processor.

\section{Experimental setup}
\label{sec:setup}

\subsection{Speech processor}

\begin{figure}
  \centering
  \includegraphics{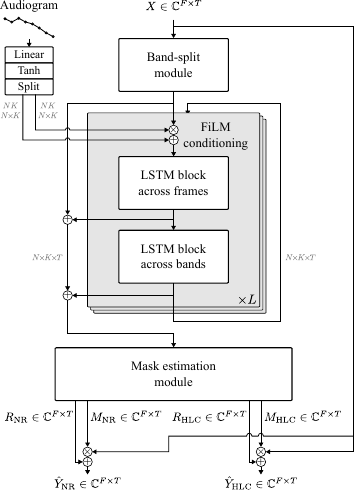}
  \caption{Speech processor based on \gls{bsrnn}~\cite{luo2023music}. Features extracted from the audiogram are fed to each layer using FiLM conditioning. The mask estimation module outputs a complex-valued mask and residual spectrogram for the denoised signal, the compensated signal, or both.}
  \label{fig:bsrnn}
  \vspace{-1pt}
\end{figure}

The speech processor is based on the \gls{bsrnn}~\cite{luo2023music}.
\gls{bsrnn} achieves state-of-the-art results for speech enhancement~\cite{yu2023efficient,yu2023high}, and provides an excellent trade-off between computational complexity and performance~\cite{zhang2024beyond}.
Figure~\ref{fig:bsrnn} shows an overview of the speech processor architecture.
The speech processor takes as input the \gls{stft} of the noisy speech $X \in \mathbb{C}^{F \times T}$, where $F$ is the number of frequency bins and $T$ is the number of frames.
The band-split module transforms the real and imaginary part of $K$ pre-defined frequency bands into a fixed number of channels $N$ using band-specific fully connected layers.
Dual-path modelling~\cite{luo2020dual} across time frames and frequency bands is performed using residual \gls{lstm} blocks in $L$ layers.
Features are extracted from the audiogram using a fully connected layer followed by Tanh activation, and fed to each layer using FiLM conditioning~\cite{perez2018film}.
The mask estimation module predicts both a mask and a residual spectrogram similarly to~\cite{yu2023high} for each output signal using band-specific \glspl{mlp}.
The mask and residual spectrogram for the denoised signal are denoted as $M_\mathrm{NR} \in \mathbb{C}^{F \times T}$ and $R_\mathrm{NR} \in \mathbb{C}^{F \times T}$, respectively.
The mask and residual spectrogram for the compensated signal are denoted as $M_\mathrm{HLC} \in \mathbb{C}^{F \times T}$ and $R_\mathrm{HLC} \in \mathbb{C}^{F \times T}$, respectively.
The \gls{stft} of the denoised and compensated signals $\hat{Y}_\mathrm{NR}$ and $\hat{Y}_\mathrm{HLC}$ are calculated as
\begin{align}
  \hat{Y}_\mathrm{NR} &= M_\mathrm{NR} \odot X + R_\mathrm{NR}, \\
  \hat{Y}_\mathrm{HLC} &= M_\mathrm{HLC} \odot X + R_\mathrm{HLC}.
\end{align}

The \gls{stft} uses a frame length of \qty{32}{\milli\second}, a hop size of \qty{16}{\milli\second}, and a Hann window.
We use ${K=27}$ frequency bands with a bandwidth of \qty{200}{\hertz} between \qtylist{0;4}{\kilo\hertz}, \qty{500}{\hertz} between \qtylist{4;7}{\kilo\hertz}, and \qty{1}{\kilo\hertz} between \qtylist{7;8}{\kilo\hertz}.
The number of channels is ${N=64}$, and the number of layers is ${L=6}$.
The number of trainable parameters varies between \qtylist{4.1;4.7}{\mega\nothing} depending on whether the speech processor is trained for \gls{nr}, \gls{hlc}, or both.
Processing a 1-\unit{\second}-long input requires \num{4.6}~\gls{gmac} operations.

\subsection{Differentiable auditory model}

The differentiable auditory model is based on the \gls{casp}~\cite{jepsen2008computational}.
\gls{casp} can successfully predict the behavior of \gls{nh} listeners in a wide range of psychoacoustic tasks, such as intensity discrimination, spectral and temporal masking, and modulation detection.
It can also simulate hearing loss, and predict the average behavior of \gls{hi} listeners~\cite{jepsen2011characterizing}.
The original model includes outer and middle ear filters, a \gls{drnl} filterbank~\cite{lopez2001human}, an inner hair cell transduction stage, adaptation loops~\cite{dau1996quantitative}, and a modulation filterbank~\cite{dau1997modeling}.

A simplified and differentiable version of \gls{casp} is implemented.
The outer ear filter is removed.
The gammatone and low-pass filters in the \gls{drnl} filterbank are implemented as 32-\unit{\milli\second}-long \gls{fir} filters to reduce training runtimes.
We use 31 filters with center frequencies spaced by one unit on the \gls{erb}-rate scale between \qtylist{80;7643}{\hertz}.
The inner hair cell transduction stage consists of half-wave rectification and low-pass filtering with a cutoff frequency of \qty{1}{\kilo\hertz}.
The adaptation loops are replaced with an instantaneous log-compression stage achieving the same steady-state gain.
The modulation filterbank is removed.

Following previous studies~\cite{lopez2012behavioral,zaar2022predicting,relano2023evaluating}, outer and inner hair cell hearing losses $\mathrm{HL}_\mathrm{OHC}$ and $\mathrm{HL}_\mathrm{IHC}$ are defined for each frequency in \unit{\decibel} as
\begin{align}
  \mathrm{HL}_\mathrm{OHC} &= \min \biggl( \frac{2}{3} \mathrm{HL}_\mathrm{tot}, \mathrm{HL}_\mathrm{OHC}^\mathrm{max} \biggr), \\
  \mathrm{HL}_\mathrm{IHC} &= \mathrm{HL}_\mathrm{tot} - \mathrm{HL}_\mathrm{OHC},
\end{align}
where $\mathrm{HL}_\mathrm{tot}$ is the total hearing loss as indicated by the audiogram, and $\mathrm{HL}_\mathrm{OHC}^\mathrm{max}$ is the maximum outer hair cell loss that can be modeled by the \gls{drnl} filterbank.
Outer and inner hair cell losses are applied as gain functions to the broken stick non-linearity of the \gls{drnl} filterbank and to the output of the inner hair cell transduction stage, respectively.
Audiogram thresholds are extrapolated to the center frequencies of the \gls{drnl} filterbank using linear interpolation.

\subsection{Datasets}

The speech processors are trained with noisy and reverberant speech generated using simulated \glspl{rir}.
The clean speech utterances are selected from DNS5~\cite{dubey2024icassp}, LibriSpeech~\cite{panayotov2015librispeech}, MLS~\cite{pratap2020mls}, VCTK~\cite{veaux2013voice}, and EARS~\cite{richter2024ears}.
The noise segments are selected from DNS5~\cite{dubey2024icassp}, WHAM!~\cite{wichern2019wham}, FSD50K~\cite{fonseca2022fsd50k}, and FMA~\cite{defferrard2017fma}.
The total amount of available speech and noise is \qtylist{1713;541}{\hour}, respectively.
The \glspl{rir} are simulated as in~\cite{luo2024fast}.
Each scene is simulated by placing one speech source and up to three noise sources in the same room.
The room size is randomly selected between \qtyproduct{3x3x2.5}{} and \qtyproduct{10x10x4}{\cubic\metre}.
The reverberation time $T_{60}$ is randomly selected between \qtylist{0.1;0.7}{\second}.
The \gls{snr} between the reverberant speech and each reverberant noise source is randomly selected between \qtylist{-10;20}{\decibel}.
Early reflections are included in the clean signal $y$ using a reflection boundary of \qty{50}{\milli\second}~\cite{roman2013speech}.
Scenes are generated on-the-fly during training.
For testing, \num{1000} scenes are generated using speech utterances from Clarity~\cite{cox2022clarity}, noise segments from TUT~\cite{mesaros2016tut}, and \glspl{rir} from DNS5~\cite{dubey2024icassp}.
All the considered datasets are publicly available.
The sampling frequency is \qty{16}{\kilo\hertz}.

\subsection{Training}

The speech processors are trained with \num{2000000} 4-\unit{\second}-long scenes.
We use a batch size of 32 and the Adam optimizer~\cite{kingma2015adam} with an initial learning rate of $1e^{-3}$.
The learning rate is reduced by a factor of 0.99 every \num{10000} scenes.
Gradients are clipped with a maximum $L_2$~norm of 5.
Training takes approximately \qty{36}{\hour} on a single A100 \qty{40}{\giga\byte} \gls{gpu}.

\subsection{Audiograms}

We consider the 10 standard audiograms from~\cite{bisgaard2010standard} as well as the \gls{nh} audiogram.
For each training scene, an audiogram $a$ is randomly selected, and a random jitter in [-10,\,10]~\unit{\decibel} is applied to each threshold to increase diversity.
Thresholds are finally clipped to [0,\,105]~\unit{\decibel}.
During testing, each scene is processed for each of the 11 profiles.
Audiogram frequencies are fixed to \qtylist{250;375;500;750;1000;1500;2000;3000;4000;6000}{\hertz}.

\subsection{Configurations}

The following speech processor configurations are compared:
\begin{itemize}
  \item \textit{BSRNN-SDR}: system trained for \gls{nr}-only using \gls{sdr}.
  \item \textit{BSRNN-NR}: system trained for \gls{nr}-only using Eq.~\eqref{eq:loss_nr}.
  \item \textit{BSRNN-HLC}: system trained for \gls{hlc}-only using Eq.~\eqref{eq:loss_hlc}.
  \item \textit{BSRNN-NR-HLC}: system trained for uncontrollable joint \gls{nr} and \gls{hlc} using Eq.~\eqref{eq:loss_nr_hlc}.
  \item \textit{BSRNN-C-NR-HLC}: system trained for controllable joint \gls{nr} and \gls{hlc} using Eq.~\eqref{eq:loss_c_nr_hlc}.
\end{itemize}
Additionally, the systems using auditory model-based objectives are trained with either the \gls{mse} or the \gls{mae} as the loss function $\ell$.

\section{Results}
\label{sec:results}

\begin{table}
  \scriptsize
  \setlength{\tabcolsep}{2pt}
  \caption{Objective metrics for the different speech processor configurations.}
  \label{tab:results}
  \centering
  \begin{tabular}{@{}lcccSSSSS@{}}
    \toprule
    & $\ell$ & $\alpha$ & $a$ & {SDR} & {PESQ} & {ESTOI} & {HASPI} & {HASQI} \\
    \midrule
    Noisy & - & - & NH & -0.15 & 1.31 & 0.58 & 0.85 & 0.33 \\
    BSRNN-SDR & - & - & NH & \bfseries 13.21 & \bfseries 2.19 & 0.85 & \bfseries 0.96 & 0.47 \\
    BSRNN-NR & MSE & - & NH & 11.78 & 1.80 & 0.82 & 0.93 & 0.42 \\
    BSRNN-NR & MAE & - & NH & 12.80 & 1.96 & \bfseries 0.85 & 0.95 & 0.46 \\
    BSRNN-HLC & MSE & - & NH & -0.25 & 1.31 & 0.58 & 0.85 & 0.33 \\
    BSRNN-HLC & MAE & - & NH & -0.12 & 1.31 & 0.58 & 0.85 & 0.33 \\
    BSRNN-NR-HLC & MSE & - & NH & 9.45 & 1.67 & 0.77 & 0.91 & 0.38 \\
    BSRNN-NR-HLC & MAE & - & NH & 7.16 & 1.16 & 0.84 & 0.89 & 0.37 \\
    BSRNN-C-NR-HLC & MSE & 0.0 & NH & -0.20 & 1.31 & 0.58 & 0.85 & 0.33 \\
    BSRNN-C-NR-HLC & MAE & 0.0 & NH & -0.17 & 1.31 & 0.58 & 0.86 & 0.33 \\
    BSRNN-C-NR-HLC & MSE & 0.8 & NH & 9.10 & 1.64 & 0.68 & 0.87 & 0.37 \\
    BSRNN-C-NR-HLC & MAE & 0.8 & NH & 9.91 & 1.82 & 0.70 & 0.92 & 0.44 \\
    BSRNN-C-NR-HLC & MSE & 1.0 & NH & 10.88 & 1.76 & 0.80 & 0.93 & 0.41 \\
    BSRNN-C-NR-HLC & MAE & 1.0 & NH & 13.00 & 2.14 & 0.85 & 0.95 & \bfseries 0.48 \\
    \midrule
    Noisy & - & - & HI & -0.15 & 1.31 & 0.58 & 0.39 & 0.23 \\
    BSRNN-SDR & - & - & HI & \bfseries 13.21 & \bfseries 2.19 & 0.85 & 0.43 & 0.25 \\
    BSRNN-NR & MSE & - & HI & 11.78 & 1.80 & 0.82 & 0.41 & 0.25 \\
    BSRNN-NR & MAE & - & HI & 12.80 & 1.96 & \bfseries 0.85 & 0.43 & 0.26 \\
    BSRNN-HLC & MSE & - & HI & -38.38 & 1.05 & 0.45 & 0.64 & 0.27 \\
    BSRNN-HLC & MAE & - & HI & -39.82 & 1.05 & 0.47 & 0.68 & 0.27 \\
    BSRNN-NR-HLC & MSE & - & HI & -34.29 & 1.07 & 0.59 & 0.74 & 0.32 \\
    BSRNN-NR-HLC & MAE & - & HI & -35.17 & 1.11 & 0.65 & \bfseries 0.80 & 0.33 \\
    BSRNN-C-NR-HLC & MSE & 0.0 & HI & -38.24 & 1.06 & 0.47 & 0.64 & 0.28 \\
    BSRNN-C-NR-HLC & MAE & 0.0 & HI & -39.94 & 1.06 & 0.48 & 0.67 & 0.28 \\
    BSRNN-C-NR-HLC & MSE & 0.8 & HI & -24.60 & 1.08 & 0.51 & 0.65 & 0.36 \\
    BSRNN-C-NR-HLC & MAE & 0.8 & HI & -26.29 & 1.09 & 0.51 & 0.68 & \bfseries 0.36 \\
    BSRNN-C-NR-HLC & MSE & 1.0 & HI & 10.97 & 1.76 & 0.80 & 0.41 & 0.24 \\
    BSRNN-C-NR-HLC & MAE & 1.0 & HI & 13.04 & 2.14 & 0.85 & 0.43 & 0.26 \\
    \bottomrule
  \end{tabular}
\end{table}

The different speech processor configurations are evaluated using \gls{sdr}, \gls{pesq}~\cite{rix2001perceptual}, \gls{estoi}~\cite{jensen2016algorithm}, \gls{haspi}~\cite{kates2014hearing}, and \gls{hasqi}~\cite{kates2010hearing}.
\gls{sdr}, \gls{pesq}, and \gls{estoi} reflect \gls{nr} performance, while \gls{haspi} and \gls{hasqi} reflect joint \gls{nr} and \gls{hlc} performance.
The results are averaged separately for scenes processed with \gls{nh} and \gls{hi} audiograms $a$, and reported in Tab.~\ref{tab:results}.
Systems trained for joint \gls{nr} and \gls{hlc} can be prompted with a \gls{nh} audiogram to compare them with systems trained for \gls{nr}-only.
When prompted with \gls{hi} audiograms, systems trained for \gls{nr}-only achieve poor \gls{haspi} and \gls{hasqi} results, since \gls{hlc} is required.
Conversely, systems trained for \gls{hlc} achieve poor \gls{sdr}, \gls{pesq}, and \gls{estoi} results, since the provided amplification causes the output to deviate from the clean signal.
For the controllable system BSRNN-C-NR-HLC trained for joint \gls{nr} and \gls{hlc}, setting the mixing parameter ${\alpha=0}$ enables \gls{hlc}-only, while ${\alpha=1}$ enables \gls{nr}-only.
Key observations include:
\begin{itemize}
  \item When prompted with ${\alpha=1}$, the controllable system BSRNN-C-NR-HLC trained with \gls{mae} outperforms its counterpart BSRNN-NR trained for \gls{nr}-only in terms of \gls{sdr} and \gls{pesq}.
  This is observed despite BSRNN-C-NR-HLC being designed to predict both a denoised and a compensated signal for a wide range of listener profiles using a similar number of trainable parameters.
  This aligns with the assumption that learning multiple related tasks simultaneously can produce better internal representations and stronger generalization compared to learning each task separately~\cite{baxter2000model}.
  \item Similarly, BSRNN-C-NR-HLC achieves superior \gls{hasqi} results to the uncontrollable system BSRNN-NR-HLC for \gls{hi} listeners and ${\alpha=0.8}$, and superior \gls{sdr} and \gls{pesq} results for \gls{nh} listeners and ${\alpha=1}$, even though BSRNN-C-NR-HLC predicts both a denoised and a compensated signal using a similar number of trainable parameters.
  \item While the highest \gls{sdr} and \gls{pesq} results are achieved by BSRNN-SDR, systems trained using the proposed differentiable auditory model achieve comparable \gls{nr} performance.
  This suggests that the output of the proposed differentiable auditory model is a valid training target for \gls{nr}.
\end{itemize}

\begin{figure}
  \centering
  \includegraphics{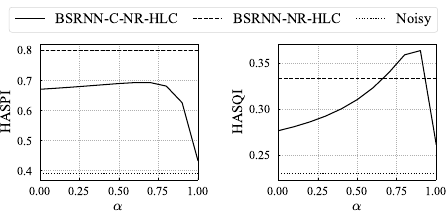}
  \caption{\gls{haspi} and \gls{hasqi} as a function of the mixing parameter $\alpha$, when using \gls{hi} audiograms and \gls{mae} loss.}
  \label{fig:alphaplot}
\end{figure}

Figure~\ref{fig:alphaplot} shows the \gls{haspi} and \gls{hasqi} results as a function of the mixing parameter $\alpha$ for the controllable system BSRNN-C-NR-HLC, using \gls{hi} audiograms and \gls{mae} loss.
The results for the uncontrollable system BSRNN-NR-HLC and the noisy input are plotted as horizontal lines.
BSRNN-C-NR-HLC achieves optimal \gls{haspi} and \gls{hasqi} results for ${\alpha=0.7}$ and ${\alpha=0.9}$, respectively.
This suggests that a balanced combination of \gls{nr} and \gls{hlc} provides optimal speech intelligibility and perceptual quality for hearing aid users.
While BSRNN-C-NR-HLC does not outperform BSRNN-NR-HLC in terms of \gls{haspi} for any ${\alpha}$ value, it does so in terms of \gls{hasqi} for ${\alpha \in [0.7, 0.9]}$.

\section{Conclusion}

We proposed a novel framework for training a system capable of controllable joint \gls{nr} and \gls{hlc}.
This approach eliminates the need for training auditory model emulators, training for individual listener profiles, or training for a fixed balance between \gls{nr} and \gls{hlc}.
The system demonstrates comparable performance to specialized systems designed for either \gls{nr}-only or \gls{hlc}-only, as measured by objective metrics.
Additionally, the ability to adjust the balance between \gls{nr} and \gls{hlc} enables optimal \gls{haspi} and \gls{hasqi} results, underscoring its relevance for hearing aid users.
Future work will evaluate the system with listening tests, investigate the influence of each differentiable auditory model stage on performance, and investigate the benefit of adjusting ${\alpha}$ dynamically based on short-time acoustic features.

\bibliographystyle{bibtex/IEEEtran}
\bibliography{bibtex/IEEEabrv, bibtex/abbrv, bibtex/mybib}

\end{document}